\documentclass[9pt,conference]{IEEEtran}

\usepackage{waspaa25} 

\usepackage{bm} 

\title{Robust Speech Activity Detection in the Presence of Singing Voice}

\name{Philipp Grundhuber$^{1}$,
      Mhd Modar Halimeh$^{1}$,
      Martin Strauß$^{1}$,
      Emanu\"el A. P. Habets$^{2}$}
\address{$^{1}$Fraunhofer Institute for Integrated Circuits IIS, Erlangen, Germany \;
$^{2}$International Audio Laboratories Erlangen, Germany\\
}

\usepackage{comment}
\usepackage{tikz}
\usepackage{acronym}
\usepackage{multirow}
\usepackage{pbalance}
\usepackage{subfigure}

\acrodef{ACC}[ACC]{Accuracy}
\acrodef{AUC}[AUC]{Area Under the Curve}
\acrodef{AUC-ROC}[AUC-ROC]{Area under curve - receiver operating characteristic}
\acrodef{CE}[CE]{Cross Entropy}
\acrodef{CNN}[CNN]{Convolutional Neural Network}
\acrodef{DE}[DE]{Dialogue Enhancement}
\acrodef{DS}[DS]{Dialogue Separation}
\acrodef{FPR}[FPR]{False Positive Rate}
\acrodef{MACs}[MACs]{Multiply-Accumulate Operations}
\acrodef{SAD}[SAD]{Speech Activity Detection}
\acrodef{SNR}[SNR]{Signal-to-Noise Ratio}
\acrodef{SR-SAD}[SR-SAD]{Singing-Robust Speech Activity Detection}
\acrodef{SR-SAD-LC}[SR-SAD-LC]{SR-SAD Low Complexity}
\acrodef{MFCC}[MFCC]{Mel-Frequency Cepstral Coefficient}
\acrodef{LSTM}{Long Short-Term Memory}
\acrodef{TCNN}[TCNN]{Temporal Convolutional Neural Network}
\acrodef{LKFS}[LKFS]{Loudness, K-weighted, relative to full scale}
\acrodef{RTF}[RTF]{Real-Time Factor}
\acrodef{ROC}[ROC]{Receiver Operating Characteristic}
\acrodef{TPR}[TPR]{True Positive Rate}
\acrodef{TNR}[TNR]{True Negative Rate}
\acrodef{SiRR}[SiRR]{Singing Rejection Rate}
\acrodef{SR}[SR]{Singing Robust}
\acrodef{GRU}[GRU]{Gated Recurrent Unit}
\acrodef{LRCN}[LRCN]{Long-term Recurrent Convolutional Network}
\acrodef{STA-VAD}{Spectro-Temporal Attention-Based VAD}
\acrodef{RNN}{Recurrent Neural Network}

\begin{document}

\maketitle

\begin{abstract}
\Ac{SAD} systems often misclassify singing as speech, leading to degraded performance in applications such as dialogue enhancement and automatic speech recognition. We introduce \ac{SR-SAD}, a neural network designed to robustly detect speech in the presence of singing. Our key contributions are: i)~a training strategy using controlled ratios of speech and singing samples to improve discrimination, ii)~a computationally efficient model that maintains robust performance while reducing inference runtime, and iii)~a new evaluation metric tailored to assess \ac{SAD} robustness in mixed speech-singing scenarios. Experiments on a challenging dataset spanning multiple musical genres show that SR-SAD maintains high speech detection accuracy (AUC = 0.919) while rejecting singing. By explicitly learning to distinguish between speech and singing, SR-SAD enables more reliable \ac{SAD} in mixed speech-singing scenarios.
\end{abstract}

\section{Introduction}

\Acf{SAD}, also known as voice activity detection, is essential for isolating speech segments in audio signals. It underpins various speech-processing applications, ensuring accurate speech extraction from mixed audio sources. However, distinguishing speech from other vocalizations, such as singing, remains a challenge. Existing \ac{SAD} systems often misclassify singing as speech, leading to errors in downstream processing \cite{Thompson2014}. Our primary research objective is to improve the classification accuracy of general-purpose SAD systems in the presence of singing.

Research in this domain can be broadly categorized into two complementary approaches: i)~joint processing and ii)~distinct voice classification. Joint processing approaches include multi-event audio classification techniques like YAMNET \cite{yamnet}, which enable fine-tuning for use cases like speech detection in the presence of singing. In a different study \cite{bai2024jointly}, separating distinct stems for speech, singing, and music directly from a stereo mixture improved singing-robust automatic speech recognition for media with mixed speech and singing. Sarasola et al. \cite{sarasola2019application} presented speech and singing classification using only pitch-derived parameters. Their work demonstrated that pitch stability and musical note characteristics can provide discrimination between speech and singing, even with short audio segments and across different languages and singing styles.

However, most research addresses speech processing and singing processing as separate domains \cite{sharma2022comprehensive}. \ac{SAD} systems are utilized as a pre-processing step to detect speech in, e.g., personalized \ac{SAD} technologies such as ECAPA-TDNN \cite{desplanques2020ecapatdnn}, NeXt-TDNN \cite{Heo_2024}, and AS-pVAD \cite{liu2024pvad} that demonstrate significant progress in speaker identification. Concurrently, researchers are addressing challenges in low \ac{SNR} environments, as exemplified by the STA-VAD approach \cite{lee2019spectro}.

In the field of singing voice detection, conventional approaches using \acp{MFCC} are often challenged by false positives, particularly when instruments produce pitch-continuous sounds similar to vocals. To address this, \cite{lehner2014reduction} proposed a set of alternative features achieving improved discrimination between vocals and instrumental sounds. Schl\"uter \cite{schluter2015exploring} further advanced the field by employing \acp{CNN} with data augmentation techniques like pitch shifting and time stretching. Singing diarization is extended in \cite{9767568} to detect multiple voices. Further, Zhang et al. \cite{electronics9091458} developed a singing voice detection system using a  Long-term Recurrent Convolutional Network that combines \ac{CNN} and Long Short-Term Memory capabilities, incorporating singing voice separation preprocessing and temporal postprocessing. While effective for music analysis, these methods focus on separating vocals from accompaniment rather than distinguishing singing from speech.

The impact of singing voice on speech detection mechanisms remains underexplored \cite{sharma2022comprehensive} despite its importance in, e.g., robust dialogue enhancement systems. Differences between singing and speech — such as higher fundamental frequency \cite{10.1121/10.0001526}, lower temporal rate, and increased pitch stability \cite{doi:10.1126/sciadv.adm9797, 9829265} — suggest the potential for distinct classification \cite{sarasola2019application}. However, it is plausible to assume similar characteristics for emotional speech, laughter, crying, and other vocal expressions \cite{9383000}. Due to these acoustic similarities, clear classification boundaries between speech and singing must be established.

In this paper, we present \ac{SR-SAD}. Our key contributions are: i)~a training strategy using controlled ratios of speech and singing samples to improve discrimination, ii)~a computationally efficient model that preserves performance while reducing inference runtime, and iii)~a novel evaluation metric tailored to assess \ac{SAD} robustness in mixed speech-singing scenarios.

\section{Problem Formulation}
Given a monaural input audio mixture $x[n]$ sampled at a sampling frequency $f_\textrm{s}$, we aim to develop a binary classification function $f(\cdot)$ that labels time frames as either containing speech or not. Let $\mathbf{x}_i \in \mathbb{R}^N$ represent the $i$-th frame of length $N$ samples. The function $f: \mathbb{R}^N \rightarrow \{0,1\}$ maps each frame to a binary label:
\begin{equation}
    f(\mathbf{x}_i) = \begin{cases} 
    1 & \text{if $\mathbf{x}_i$ contains speech}; \\
    0 & \text{otherwise}.
    \end{cases}
\end{equation}
Hence, the output equals $1$ for frames containing speech or speech and singing simultaneously. In contrast, the output equals $0$ for all speech-free frames, including singing voice, instrumental music, ambient sounds, and silence. The system must maintain robustness across diverse acoustic conditions, including different speakers, various musical styles, and varying \ac{SNR} levels.

\section{Proposed Speech Activity Detection}
\subsection{Architectures}

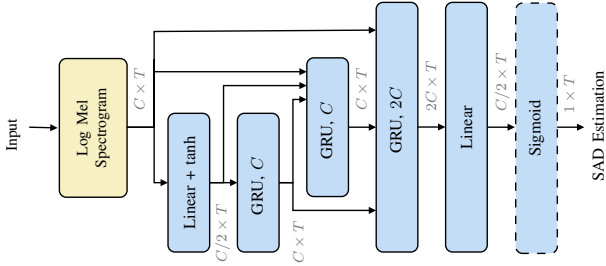
\begin{figure}[t]
   \centering
   \vspace{-2.5cm}
   \rotatebox{90}{ 
     \scalebox{0.7}{\parbox{\linewidth}{\tikzset{every picture/.style={line width=0.75pt}} 

\begin{tikzpicture}[x=0.75pt,y=0.75pt,yscale=-1,xscale=1]

\draw  [fill={rgb, 255:red, 166; green, 205; blue, 243 }  ,fill opacity=0.67 ] (30.5,145) .. controls (30.5,142.24) and (32.74,140) .. (35.5,140) -- (125.5,140) .. controls (128.26,140) and (130.5,142.24) .. (130.5,145) -- (130.5,165) .. controls (130.5,167.76) and (128.26,170) .. (125.5,170) -- (35.5,170) .. controls (32.74,170) and (30.5,167.76) .. (30.5,165) -- cycle ;
\draw  [fill={rgb, 255:red, 166; green, 205; blue, 243 }  ,fill opacity=0.67 ] (30.5,195) .. controls (30.5,192.24) and (32.74,190) .. (35.5,190) -- (125.5,190) .. controls (128.26,190) and (130.5,192.24) .. (130.5,195) -- (130.5,215) .. controls (130.5,217.76) and (128.26,220) .. (125.5,220) -- (35.5,220) .. controls (32.74,220) and (30.5,217.76) .. (30.5,215) -- cycle ;
\draw  [fill={rgb, 255:red, 166; green, 205; blue, 243 }  ,fill opacity=0.67 ] (70.5,245) .. controls (70.5,242.24) and (72.74,240) .. (75.5,240) -- (165.5,240) .. controls (168.26,240) and (170.5,242.24) .. (170.5,245) -- (170.5,265) .. controls (170.5,267.76) and (168.26,270) .. (165.5,270) -- (75.5,270) .. controls (72.74,270) and (70.5,267.76) .. (70.5,265) -- cycle ;
\draw  [fill={rgb, 255:red, 166; green, 205; blue, 243 }  ,fill opacity=0.67 ] (30.5,295) .. controls (30.5,292.24) and (32.74,290) .. (35.5,290) -- (205.5,290) .. controls (208.26,290) and (210.5,292.24) .. (210.5,295) -- (210.5,315) .. controls (210.5,317.76) and (208.26,320) .. (205.5,320) -- (35.5,320) .. controls (32.74,320) and (30.5,317.76) .. (30.5,315) -- cycle ;
\draw  [fill={rgb, 255:red, 166; green, 205; blue, 243 }  ,fill opacity=0.67 ] (30.5,345) .. controls (30.5,342.24) and (32.74,340) .. (35.5,340) -- (205.5,340) .. controls (208.26,340) and (210.5,342.24) .. (210.5,345) -- (210.5,365) .. controls (210.5,367.76) and (208.26,370) .. (205.5,370) -- (35.5,370) .. controls (32.74,370) and (30.5,367.76) .. (30.5,365) -- cycle ;
\draw    (120.5,130) -- (169.5,130) -- (104.17,130) -- (80.5,130) -- (80.5,137) ;
\draw [shift={(80.5,140)}, rotate = 270] [fill={rgb, 255:red, 0; green, 0; blue, 0 }  ][line width=0.08]  [draw opacity=0] (5.36,-2.57) -- (0,0) -- (5.36,2.57) -- cycle    ;
\draw    (80.5,170) -- (80.5,180) -- (150.5,180) -- (150.5,237) ;
\draw [shift={(150.5,240)}, rotate = 270] [fill={rgb, 255:red, 0; green, 0; blue, 0 }  ][line width=0.08]  [draw opacity=0] (5.36,-2.57) -- (0,0) -- (5.36,2.57) -- cycle    ;
\draw    (120.5,110) -- (120.5,130) -- (189.5,130) -- (190.48,287) ;
\draw [shift={(190.5,290)}, rotate = 269.64] [fill={rgb, 255:red, 0; green, 0; blue, 0 }  ][line width=0.08]  [draw opacity=0] (5.36,-2.57) -- (0,0) -- (5.36,2.57) -- cycle    ;
\draw    (80.5,220) -- (80.5,230) -- (140.5,230) -- (140.5,237) ;
\draw [shift={(140.5,240)}, rotate = 270] [fill={rgb, 255:red, 0; green, 0; blue, 0 }  ][line width=0.08]  [draw opacity=0] (5.36,-2.57) -- (0,0) -- (5.36,2.57) -- cycle    ;
\draw    (80.5,220) -- (80.5,230) -- (60.5,230) -- (60.5,287) ;
\draw [shift={(60.5,290)}, rotate = 270] [fill={rgb, 255:red, 0; green, 0; blue, 0 }  ][line width=0.08]  [draw opacity=0] (5.36,-2.57) -- (0,0) -- (5.36,2.57) -- cycle    ;
\draw    (120.5,270) -- (120.5,287) ;
\draw [shift={(120.5,290)}, rotate = 270] [fill={rgb, 255:red, 0; green, 0; blue, 0 }  ][line width=0.08]  [draw opacity=0] (5.36,-2.57) -- (0,0) -- (5.36,2.57) -- cycle    ;
\draw    (120.5,370) -- (120.5,387) ;
\draw [shift={(120.5,390)}, rotate = 270] [fill={rgb, 255:red, 0; green, 0; blue, 0 }  ][line width=0.08]  [draw opacity=0] (5.36,-2.57) -- (0,0) -- (5.36,2.57) -- cycle    ;
\draw    (80.5,170) -- (80.5,180) -- (80.5,180) -- (80.5,187) ;
\draw [shift={(80.5,190)}, rotate = 270] [fill={rgb, 255:red, 0; green, 0; blue, 0 }  ][line width=0.08]  [draw opacity=0] (5.36,-2.57) -- (0,0) -- (5.36,2.57) -- cycle    ;
\draw    (120.62,319.87) -- (120.62,336.87) ;
\draw [shift={(120.62,339.87)}, rotate = 270] [fill={rgb, 255:red, 0; green, 0; blue, 0 }  ][line width=0.08]  [draw opacity=0] (5.36,-2.57) -- (0,0) -- (5.36,2.57) -- cycle    ;
\draw    (120.5,420) -- (120.5,437) ;
\draw [shift={(120.5,440)}, rotate = 270] [fill={rgb, 255:red, 0; green, 0; blue, 0 }  ][line width=0.08]  [draw opacity=0] (5.36,-2.57) -- (0,0) -- (5.36,2.57) -- cycle    ;
\draw    (120.5,110) -- (120.5,130) -- (160.5,130) -- (160.5,237) ;
\draw [shift={(160.5,240)}, rotate = 270] [fill={rgb, 255:red, 0; green, 0; blue, 0 }  ][line width=0.08]  [draw opacity=0] (5.36,-2.57) -- (0,0) -- (5.36,2.57) -- cycle    ;
\draw  [fill={rgb, 255:red, 243; green, 233; blue, 166 }  ,fill opacity=0.67 ] (70,66.72) .. controls (70,63.96) and (72.24,61.72) .. (75,61.72) -- (165,61.72) .. controls (167.76,61.72) and (170,63.96) .. (170,66.72) -- (170,105.5) .. controls (170,108.26) and (167.76,110.5) .. (165,110.5) -- (75,110.5) .. controls (72.24,110.5) and (70,108.26) .. (70,105.5) -- cycle ;
\draw    (120.5,40) -- (120.07,57.5) ;
\draw [shift={(120,60.5)}, rotate = 271.4] [fill={rgb, 255:red, 0; green, 0; blue, 0 }  ][line width=0.08]  [draw opacity=0] (5.36,-2.57) -- (0,0) -- (5.36,2.57) -- cycle    ;
\draw  [fill={rgb, 255:red, 166; green, 205; blue, 243 }  ,fill opacity=0.67 ][dash pattern={on 4.5pt off 4.5pt}] (29.5,395.5) .. controls (29.5,392.74) and (31.74,390.5) .. (34.5,390.5) -- (205,390.5) .. controls (207.76,390.5) and (210,392.74) .. (210,395.5) -- (210,415.5) .. controls (210,418.26) and (207.76,420.5) .. (205,420.5) -- (34.5,420.5) .. controls (31.74,420.5) and (29.5,418.26) .. (29.5,415.5) -- cycle ;

\draw (101.5,22) node [anchor=north west][inner sep=0.75pt]  [font=\small] [align=left] {Input};
\draw (120,86.11) node  [font=\small] [align=left] {\begin{minipage}[lt]{55.28pt}\setlength\topsep{0pt}
\begin{center}
Log Mel \\Spectrogram
\end{center}

\end{minipage}};
\draw (80.5,155.5) node   [align=left] {Linear + tanh};
\draw (80.5,205) node   [align=left] {GRU, $\displaystyle C$ };
\draw (120.5,255) node   [align=left] {GRU, $\displaystyle C$ };
\draw (125.5,305) node   [align=left] {GRU, $\displaystyle 2C$};
\draw (120.5,355) node   [align=left] {Linear};
\draw (121.5,450) node   [align=left] {SAD Estimation};
\draw (155.5,120) node  [font=\footnotesize,color={rgb, 255:red, 120; green, 120; blue, 120 }  ,opacity=1 ] [align=left] {\begin{minipage}[lt]{34.17pt}\setlength\topsep{0pt}
$\displaystyle C\times T$
\end{minipage}};
\draw (46,179.5) node  [font=\footnotesize,color={rgb, 255:red, 120; green, 120; blue, 120 }  ,opacity=1 ] [align=left] {\begin{minipage}[lt]{33.66pt}\setlength\topsep{0pt}
$\displaystyle C/2\times T$
\end{minipage}};
\draw (45,234.5) node  [font=\footnotesize,color={rgb, 255:red, 120; green, 120; blue, 120 }  ,opacity=1 ] [align=left] {\begin{minipage}[lt]{34pt}\setlength\topsep{0pt}
$\displaystyle C\times T$
\end{minipage}};
\draw (155.5,280) node  [font=\footnotesize,color={rgb, 255:red, 120; green, 120; blue, 120 }  ,opacity=1 ] [align=left] {\begin{minipage}[lt]{34.17pt}\setlength\topsep{0pt}
$\displaystyle C\times T$
\end{minipage}};
\draw (155,383) node  [font=\footnotesize,color={rgb, 255:red, 120; green, 120; blue, 120 }  ,opacity=1 ] [align=left] {\begin{minipage}[lt]{34pt}\setlength\topsep{0pt}\vspace{-3pt}
$\displaystyle C/2\times T$
\end{minipage}};
\draw (154,425.5) node  [font=\footnotesize,color={rgb, 255:red, 120; green, 120; blue, 120 }  ,opacity=1 ] [align=left] {\begin{minipage}[lt]{34pt}\setlength\topsep{0pt}\vspace{6pt}
$\displaystyle 1\times T$
\end{minipage}};
\draw (155,330) node  [font=\footnotesize,color={rgb, 255:red, 120; green, 120; blue, 120 }  ,opacity=1 ] [align=left] {\begin{minipage}[lt]{34pt}\setlength\topsep{0pt}
$\displaystyle 2C\times T$
\end{minipage}};
\draw (119.75,405.5) node  [font=\normalsize] [align=left] {Sigmoid};

\end{tikzpicture}}}
   }
   \caption{\textit{Architecture for proposed SR-SAD}}
   \label{fig:SR-SAD}
\end{figure}

\begin{figure}[t]
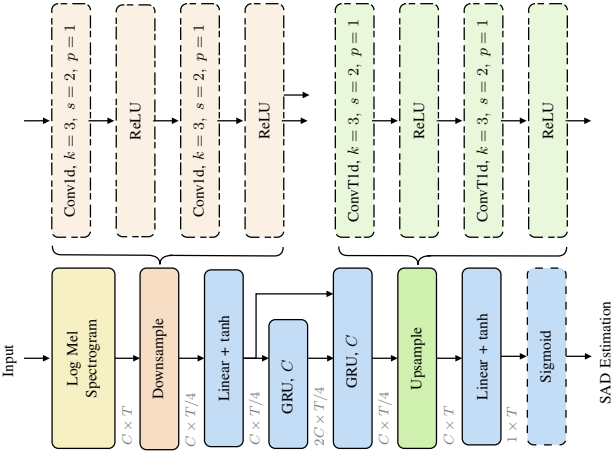

   \centering
   \vspace{0.5cm}
   \rotatebox{90}{ 
     \scalebox{0.65}{\parbox{\linewidth}{\include{SRVAD_architecture_efficient}}}
   }
   \caption{\textit{Architecture for proposed SR-SAD-LC}}
   \label{fig:SR-SAD-LC}
\end{figure}
We propose two \ac{SR-SAD} variants based on the architecture used for denoising in \cite{valin2018hybrid}. The input for both models is a $C$-dimensional mel-spectrogram with $C=80$ and $T$ time steps, computed with $N_\textrm{FFT} = 512$, a window length of $512$, and a hop length of $256$. The first model, depicted in Figure~\ref{fig:SR-SAD}, processes the spectrogram through a linear layer and a tanh activation function, followed by three cascaded bidirectional \acp{GRU}, each comprising two layers. Each \ac{GRU} processes outputs of previous \acp{GRU} concatenated with original inputs, incorporating skip-connections for frame-level prediction.

\ac{SR-SAD-LC}, shown in Figure~\ref{fig:SR-SAD-LC}, reduces computational complexity by downsampling the input along the time axis to the \ac{GRU}, reducing parameter count and inference times. Here, the downsampling is realized using strided convolutional layers. At the output of the last \ac{GRU}, the feature map is upsampled to restore the input dimensionality utilizing transposed convolutional layers. In contrast to SR-SAD, SR-SAD-LC uses a single-layer bidirectional \ac{GRU}.

\subsection{Training Dataset} \label{ssec:training}
The training data consists of four datasets:
\begin{itemize}
    \item \textsc{Speech}: DNS Challenge 2021 \cite{reddy2021interspeech}: emotional\_speech, french\_data, german\_speech, italian\_speech, read\_speech
    \item \textsc{Music} and \textsc{Singing}: MUSDB18-HQ \cite{MUSDB18HQ}, where songs containing rap or speech were removed.
    \item \textsc{Noise}: DNS Challenge 2021 \cite{reddy2021interspeech}: noise.
\end{itemize}
These datasets were randomly partitioned into training (80\%) and validation (20\%) sets, with no overlapping speakers or artists between splits. This split remained consistent across all experiments to ensure reproducibility and a fair comparison. All data was resampled to \SI{16}{kHz}. 
Training samples were created dynamically by superimposing individual speech, music, and noise signals. For each training sample, a continuous audio segment of length $T = L \cdot f_\textrm{s}$ samples is randomly selected from either the \textsc{Speech} or \textsc{Singing} dataset, where $L$ is the input chunk length in seconds and $f_\textrm{s}$ is the sampling frequency. The probability $p_\textrm{s}$ determines whether speech or singing is selected. When speech is selected, it is mixed with random noise from the \textsc{Noise} dataset. When singing is selected, it is mixed with its corresponding instrumental music from MUSDB18-HQ. The use of separate stems for singing and instrumental music allows for independent control of the singing-to-music ratio.

\subsection{Test Dataset}
The test dataset is constructed to contain speech, singing, and other background noise. Speech is sourced from the EARS dataset \cite{richter2024ears}, while all samples labeled as 'melodic' or 'vegetative' are excluded, as these do not contain speech. Music and singing samples are taken from the mixed songs of MoisesDB \cite{pereira2023moisesdb} without gain modification between the sources, and noise is taken from WHAM! \cite{Wichern2019WHAM}, which contains babble noises. The speech-like babble noises make this a highly challenging dataset. All speakers and acoustic conditions are different from training and evaluation data. All data was resampled to \SI{16}{kHz}. The length of the dataset is \SI{100}{h} with 24000 individual signals each \SI{15}{s} long. The \ac{SNR} between speech and all other signals, as well as loudness, are both uniformly sampled, with SNR ranging from \SI{-5}{dB} to \SI{10}{dB} and loudness from \SI{-30}{LKFS} to \SI{-10}{LKFS}. Speech labels are generated based on the energy of the clean speech signal. If speech is loaded into a sample, a \SI{12}{\second} excerpt is positioned randomly within the sample. This test set is used for the performance evaluation. It contains 34.7\% speech, 23.4\% singing, and 8.1\% overlapping passages of speech and singing.

\begin{table}[t]
  \caption{Audio Signal Augmentation Parameters}
  \label{tab:Augmentation}
  \centering
  \begin{tabular}{l l}
    \toprule
    \multicolumn{1}{l}{\textbf{Augmentation Method}} & \multicolumn{1}{l}{\textbf{Probability}} \\
    \midrule
         \ac{SNR} Adjustment (±7 dB range) & $80 \%$ \\
         Frequency Band Rejection (100-4000 Hz) & $80 \%$ \\
         High-Pass Filtering (cutoff: 500-4000 Hz) & $30\%$\\
         Low-Pass Filtering (cutoff: 3000-8000 Hz) & $10\%$\\
         Audio Signal Clipping & $10\%$\\
         Amplitude Scaling (0.1 to 1.0) & $40\%$\\
         White Noise Addition & $10\%$\\
         Stereo to Mono Conversion  & $100\%$\\
    \bottomrule
  \end{tabular}

\end{table}

\subsection{Training}
The mixed audio signals are processed through the augmentation scheme given in Table~\ref{tab:Augmentation}. The augmentation schemes follow common techniques \cite{ALEX2023102949}, while specific parameters were determined empirically. Ground truth labels for speech were generated using energy-based thresholding of the clean speech signal with gap filling applied to bridge silent periods shorter than \SI{300}{ms}. For singing, all labels are set to zero. An epoch consists of 100,000 audio/label pairs for training and 1,000 pairs for validation. 

Both model variants are trained on one NVIDIA GeForce RTX 2080 TI using the binary cross-entropy loss function and learning rate scheduling, which halves the learning rate when the validation loss does not reduce after 20 epochs. The ADAM optimizer \cite{kingma2014adam} was used with an initial learning rate of 0.001 and weight decay of 0.0001, and early stopping was used with a patience of 20. The training spanned between 60 and 200 epochs, depending on the hyperparameters.

\section{Experimental Setup}

\subsection{Baselines}
The proposed SAD model is compared to other state-of-the-art voice activity detection (VAD) models. All models were trained using identical data and loss to enable fair comparison. Three architectures are compared: AS-pVAD~\cite{liu2024pvad}, which uses Temporal-\ac{CNN} architecture followed by \acp{GRU} with hidden dimensions set to 100 and 80-dim logmel spectrogram input, STA-VAD~\cite{lee2019spectro}, which employs self-attention mechanisms with an 80-dim normalized mel spectrum.
AS-pVAD was designed for preprocessing speech within a personalized system, while STA-VAD targets SAD in low SNR conditions. In addition to the AS-pVAD and STA-VAD architectures, we include ResNet50 \cite{Gohari2024} as a baseline with a fully convolutional architecture for comparison. ResNet50, followed by fully connected layers, has been successfully employed for detecting edited audio in tasks such as auto-tuned singing voice detection, where it effectively distinguishes between natural and processed vocal signals based on spectrogram features \cite{Gohari2024}. SR-SAD, SR-SAD-LC, and AS-pVAD use chunk-based processing during inference, with input chunk length $L$ analyzed in Section~\ref{subsec:chunk_window_length}. ResNet50 is trained with a fixed $L = 10\,s$, following the original configuration in \cite{Gohari2024} optimized for spectro-temporal pattern recognition in audio processing tasks. These baselines were also trained using the data described in Section~\ref{ssec:training}.

\subsection{Performance Metrics}
\ac{SAD} classification performance is commonly measured by \ac{AUC} \cite{sharma2022comprehensive}, where the \ac{TPR} represents the proportion of actual speech instances correctly identified, and the \ac{FPR} reflects the proportion of non-speech instances that are mistakenly identified as speech. This metric is denoted in the following as $\mbox{AUC}$. While $\mbox{AUC}$ measures overall classification performance, it treats all non-speech content equally. Therefore, it does not specifically reflect the performance when singing is present, which is considered more challenging. The $\mbox{AUC}_{\textrm{SiRR}}$ metric is designed to directly measure the performance for the singing voice case by assessing the \ac{TPR} for speech detection against the \ac{SiRR}. The SiRR represents how well the system rejects singing content specifically, calculated as the \ac{TNR} for frames containing singing. This captures both aspects of our task: maintaining high speech detection accuracy (TPR) while effectively rejecting singing content (SiRR). A perfect system would achieve both TPR = 1 (detecting all speech) and SiRR = 1 (rejecting all singing) across all classification thresholds, resulting in $\mbox{AUC}_{\textrm{SiRR}} = 1$. Hence, a higher $\mbox{AUC}_{\textrm{SiRR}}$ indicates greater robustness against singing.

\section{Evaluation}

\subsection{Probability of Speech During Training}
The parameter $p_\textrm{s}$ controls the probability of selecting speech versus singing samples during training data generation. We evaluated values of $p_\textrm{s}$ from 0\% to 100\% in 10\% increments, with a fixed input chunk length of \SI{2}{\second} for SR-SAD, SR-SAD-LC, AS-pVAD, and \SI{10}{\second} for ResNet50 models.

\begin{figure}[t]
    \centering
    \includegraphics[width=0.5\textwidth]{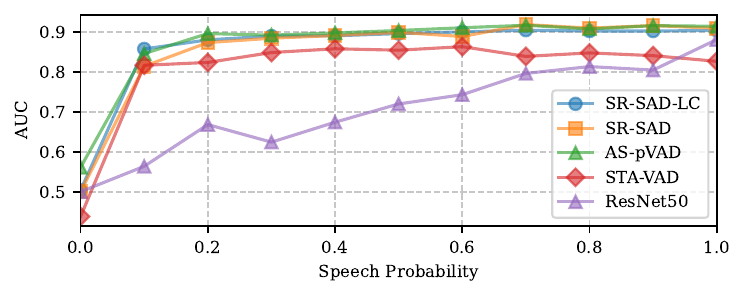}
    \caption{\textit{Impact of speech probability $p_\textrm{s}$ on model performance measured by $\mbox{$\mbox{AUC}$}$. Highest $\mbox{$\mbox{AUC}$}$ occurs at $p_s = 80\%$.}}
    \label{fig:AUC-PR_per_Speech Probability}
\end{figure}

The performance was assessed using $\mbox{AUC}$ and $\mbox{$\mbox{AUC}_{\textrm{SiRR}}$}$ metrics. As shown in Figure~\ref{fig:AUC-PR_per_Speech Probability}, SR-SAD, SR-SAD-LC, and AS-pVAD achieved peak $\mbox{AUC}$ performance of, respectively, 0.919, 0.905, and 0.917 with $p_\textrm{s}$ values of 70\%. STA-VAD, which consistently performed below the other models, achieved its peak $\mbox{AUC}$ of 0.863 at $p_\textrm{s}\!={}\!60\%$. ResNet50 showed the lowest performance among all models for most speech probabilities, but reached its peak $\mbox{AUC}$ of 0.881 at $p_\textrm{s}\!={}\!100\%$.

\begin{figure}[t]
    \centering
    \includegraphics[width=0.5\textwidth]{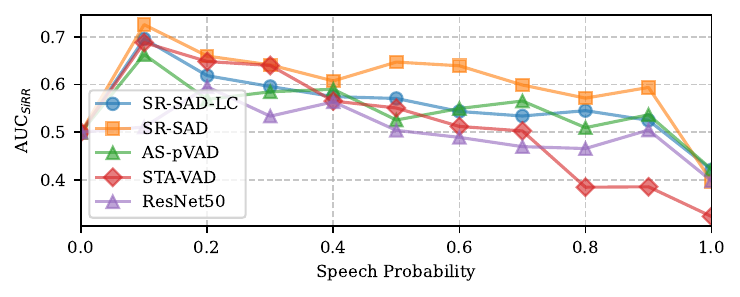}
    \caption{\textit{$\mbox{AUC}_{\textrm{SiRR}}$ across different speech probability $p_\textrm{s}$ values. Highest occurs at $p_\textrm{s}\!={}\!10\%$ and decreases as $p_\textrm{s}$ increases.}}
    \label{fig:tnr_SV_only_perCat_TNR_SV_ONLY}
\end{figure}

Analysis of $\mbox{$\mbox{AUC}_{\textrm{SiRR}}$}$ reveals an inverse relationship with $p_\textrm{s}$, as illustrated in Figure~\ref{fig:tnr_SV_only_perCat_TNR_SV_ONLY}. All models demonstrated optimal singing rejection capabilities while detecting speech correctly at $p_\textrm{s}\!={}\!10\%$, with SR-SAD achieving 0.726, SR-SAD-LC reaching 0.696, AS-pVAD attaining 0.663, and STA-VAD scoring 0.688. ResNet50 showed a different pattern, with its peak $\mbox{$\mbox{AUC}_{\textrm{SiRR}}$}$ of 0.595 occurring at $p_\textrm{s}\!={}\!20\%$. Training with speech samples only ($p_\textrm{s}\!={}\!100\%$) results in a high \ac{AUC} but minimal $\mbox{$\mbox{AUC}_{\textrm{SiRR}}$}$ performance, dropping to 0.397 for SR-SAD, 0.421 for SR-SAD-LC, 0.422 for AS-pVAD, 0.323 for STA-VAD, and 0.398 for ResNet50.

The combined analysis indicates that including at least 20\% singing samples during training substantially improves $\mbox{$\mbox{AUC}_{\textrm{SiRR}}$}$ for all models. We selected $p_\textrm{s}\!={}\!80\%$ for subsequent experiments, as this value provides competitive $\mbox{AUC}$ performance (0.910 for SR-SAD, 0.903 for SR-SAD-LC, 0.908 for AS-pVAD, and 0.848 for STA-VAD) while maintaining moderate robustness against singing voice with $\mbox{$\mbox{AUC}_{\textrm{SiRR}}$}$ values of 0.572, 0.545, 0.510, and 0.385, respectively.

\begin{figure}[!t]
    \centering
    \includegraphics[width=0.5\textwidth]{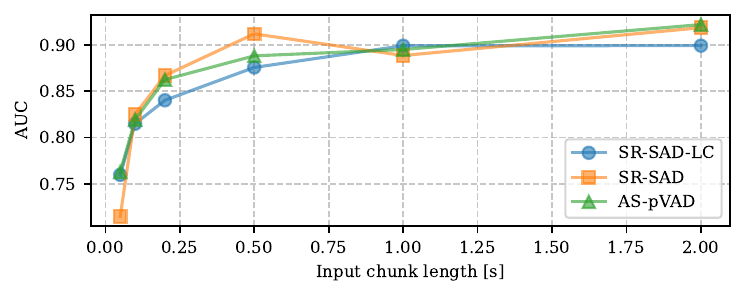}
    \caption{\textit{Effect of input chunk length $L$ on model performance measured by $\mbox{$\mbox{AUC}$}$. Longer input chunks consistently yield better performance across all architectures.}}
    \label{fig:chunk_length}
\end{figure}


\begin{figure*}[!t]
    \centering
    \includegraphics[width=\textwidth]{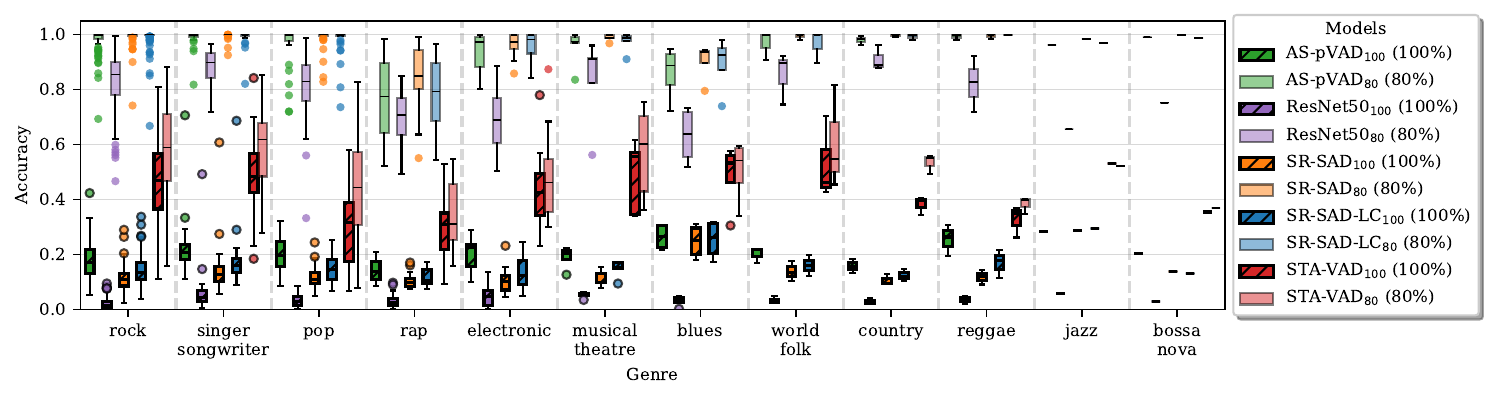}
    \caption{\textit{\ac{ACC} analysis across musical genres for models trained with $p_\textrm{s}\!={}\!80\%$ and $p_\textrm{s}\!={}\!100\%$. Higher \ac{ACC} values indicate better singing suppression.}}
    \label{fig:accuracy_box_by_genre_vocal_active_grouped_comp}
\end{figure*}

\subsection{Input Chunk Length}
\label{subsec:chunk_window_length}
We evaluated the impact of input chunk length, $L$, on model performance by utilizing chunks that are \SI{0.05}{\second}, \SI{0.1}{\second}, \SI{0.2}{\second}, \SI{0.5}{\second}, \SI{1}{\second}, and \SI{2}{\second} long for the SR-SAD, SR-SAD-LC, and AS-pVAD architectures. As shown in Figure~\ref{fig:chunk_length}, a consistent trend is observed across all architectures: shorter input chunks and their respective limited temporal context result in reduced $\mbox{AUC}$ performance. SR-SAD achieved optimal performance with \SI{2}{\second}-long chunks ($\mbox{AUC}$ = 0.919), while AS-pVAD reached its peak at \SI{2}{\second} ($\mbox{AUC}$ = 0.922). SR-SAD-LC showed no improvement of performance between \SI{1}{\second} compared to \SI{2}{\second} ($\mbox{AUC}$ = 0.899). The performance degradation was most pronounced at \SI{0.05}{\second}, where $\mbox{AUC}$ scores dropped to 0.715 for SR-SAD, 0.763 for AS-pVAD, and 0.760 for SR-SAD-LC. Although not shown here due to space constraints, analysis of $\mbox{AUC}_{\textrm{SiRR}}$ versus input chunk length showed weaker dependence on $L$ compared to $\mbox{AUC}$. Based on these findings, we selected a \SI{2}{\second} input chunk length for subsequent analyses to maximize $\mbox{AUC}$ performance across all architectures.

\subsection{Genre-Specific Performance Analysis}
We further evaluated performance on the MoisesDB singing-only dataset containing 240 songs. Since the dataset contains only one class (non-speech), AUC analysis was not applicable, and \acf{ACC} was used with standard deviation $\sigma$ across songs to indicate consistency. Songs were processed entirely, and subsequently, segments that contained singing were extracted and evaluated. The results, grouped by genre according to the dataset's metadata, are shown in Figure~\ref{fig:accuracy_box_by_genre_vocal_active_grouped_comp}. We compare all trained models with $p_\textrm{s}\!={}\!100\%$ and $p_\textrm{s}\!={}\!80\%$. Models trained with $p_\textrm{s}\!={}\!100\%$ consistently showed lower \ac{ACC} across all architectures, as detailed in Table~\ref{tab:Overall_MoisesDB}. The effect was most pronounced in ResNet50 (23.8 times higher), SR-SAD (8.05 times higher), SR-SAD-LC (6.51 times higher), and AS-pVAD (5.05 times higher), while STA-VAD showed minimal difference (1.23 times higher). All networks exhibited decreased accuracy for rap and blues genres, and lowered accuracy for electronic music. This can be attributed to rap's spoken-word characteristics and the presence of speech in electronic music intros, while blues often features soft, slow vocals which are characteristically close to speech \cite{sundberg2019acoustics}. Notably, rap music was explicitly excluded from the training set to maintain clear class definitions between speech and singing. The overall statistics in Table~\ref{tab:Overall_MoisesDB} demonstrate that SR-SAD trained with $p_\textrm{s}\!={}\!80\%$ achieved the best performance with an \ac{ACC} of 0.9786 and the lowest standard deviation of 0.0586. 
These results illustrate that including singing voice during \ac{SAD} training ($p_\textrm{s}\!={}\!80\%$) dramatically improves the model's ability to detect singing as non-speech, yielding significantly higher accuracy scores across architectures, with the proposed architecture successfully labeling singing as non-speech despite showing some limitations for rap and blues.

\subsection{Computational Complexity}
The computational requirements were evaluated using \ac{MACs}, \ac{RTF}, and the number of parameters and provided in Table~\ref{tab:computational_complexity}. All measurements were performed with an Intel(R) Xeon(R) Platinum 8260 CPU @ 2.40GHz using \SI{2}{\second} input chunks. SR-SAD-LC achieved the lowest computational cost with \SI{15.6}{M} MACs and the highest RTF of 275 while maintaining a moderate parameter count of \SI{335}{K}.

\begin{table}[!t]
    \caption{Accuracy and standard deviation $\sigma$ for MoisesDB, averaged across genres for models trained with $p_\textrm{s}\!={}\!80\%$ and $p_\textrm{s}\!={}\!100\%$.}
    \label{tab:Overall_MoisesDB}
    \centering
    \begin{tabular}{l rr rr}
    \toprule
    \textbf{Model}& \multicolumn{2}{c}{\textbf{ACC}} & \multicolumn{2}{c}{$\boldsymbol{\sigma}$} \\
    \cmidrule(lr){2-3}\cmidrule(lr){4-5}
    $p_\textrm{s}$      & $80\%$            & $100\%$& $80\%$ & $100\%$ \\
    \midrule
    $\text{SR-SAD}$     & \textbf{0.9786}   & 0.1215 & 0.0586 & 0.0632 \\
    $\text{SR-SAD-LC}$  & 0.9687            & 0.1488 & 0.0771 & 0.4589 \\
    $\text{AS-pVAD}$    & 0.9598            & 0.1902 & 0.0823 & 0.0683 \\
    $\text{STA-VAD}$    & 0.5341            & 0.4342 & 0.1708 & 0.1572 \\
    $\text{ResNet50}$   & 0.8161            & 0.0343 & 0.1175 & 0.0397 \\
    \bottomrule
    \end{tabular}
\end{table}

\begin{table}[!t]
  \caption{\ac{SAD} systems in terms of computational complexity}
  \label{tab:computational_complexity}
  \centering
  \begin{tabular}{l l l l}
    \toprule
    \multicolumn{1}{l}{\textbf{Model}} & 
                                         \multicolumn{1}{c}{\textbf{MACs}} & 
                                         \multicolumn{1}{c}{\textbf{RTF}}& 
                                         \multicolumn{1}{c}{\textbf{\# of Parameters}} \\
    \midrule
        $\text{SR-SAD-LC}$ &  \textbf{\SI{15.6}{M}} & \textbf{275}& \SI{335}{K}\\
        $\text{SR-SAD}$ & \SI{82.9}{M}  & 32& \SI{870}{K}\\
        $\text{AS-pVAD}$ & \SI{38.5}{M} & 125 & \textbf{\SI{279}{K}} \\
        $\text{STA-VAD}$ & \SI{3.35}{G}& 7& \SI{559}{K}\\
        $\text{ResNet50}$ & \SI{80.05}{G}& 15& \SI{11.4}{M}\\
    \bottomrule
  \end{tabular}
\end{table}

\subsection{Limitations}
While the proposed SR-SAD systems show promising results, several key limitations remain. The system's \SI{2}{\second} chunk-based processing prevents real-time applications in its current form. Performance analysis across musical genres revealed significant challenges with rap and blues genres. The explicit exclusion of rap music from our training dataset, while maintaining clear class definitions, likely contributed to these performance gaps. Despite improvements in the SR-SAD-LC variant, the computational requirements still exceed those of traditional SAD systems, which may limit deployment on resource-constrained devices. Additionally, for increased robustness in scenarios with overlapping speech and singing, this condition should be explicitly incorporated into the training data.

\section{Conclusion}
This paper presents SR-SAD, a novel approach to speech activity detection focused on speech-singing discrimination. Experimental results demonstrate strong performance in speech detection and singing suppression across most musical genres, with both SR-SAD variants outperforming baseline methods. Key findings include: i) the efficacy of longer input chunks and their extended temporal context for temporal pattern recognition, with performance consistently improving as chunk length increases from \SI{0.05}{\second} to \SI{2}{\second} across all architectures; ii) the critical importance of balanced training data distribution, with optimal performance achieved using 20-30\% singing examples; iii) the \ac{RNN}-based architectures (here implemented using \acp{GRU}) have superior temporal modeling capabilities compared to purely convolutional approaches (the ResNet50 baseline) for this task; and iv) the possibility of achieving substantial computational efficiency (80\% reduction) through strategic downsampling while maintaining robust performance. Future work should address the identified limitations, particularly improving performance for challenging musical genres like rap and blues, where vocal characteristics closely resemble speech, and reducing processing latency for real-time applications.


\clearpage
\bibliographystyle{IEEEtran}
\bibliography{refs25}







\end{document}